# Investigating the Influence of Visualization on Student Understanding of Quantum Superposition

Antje Kohnle, Charles Baily and Scott Ruby

*School of Physics and Astronomy, University of St Andrews, North Haugh, St Andrews, KY16 9SS, UK*

**Abstract:** Visualizations in interactive computer simulations are a powerful tool to help students develop productive mental models, particularly in the case of quantum phenomena that have no classical analogue. The QuVis Quantum Mechanics Visualization Project develops research-based interactive simulations for the learning and teaching of quantum mechanics. We describe efforts to refine the visual representation of a single-photon superposition state in the QuVis simulations. We developed various depictions of a photon incident on a beam splitter, and investigated their influence on student thinking through individual interviews. Outcomes from this study led to the incorporation of a revised visualization in all QuVis single-photon simulations. In-class trials in 2013 and 2014 using the *Interferometer Experiments* simulation in an introductory quantum physics course were used for a comparative study of the initial and revised visualizations. The class that used the revised visualization showed a lower frequency of incorrect ideas about quantum superposition, such as the photon splitting into two half-energy components.

**Keywords:** quantum mechanics, simulations, visualization, conceptual understanding.

**PACS:** 01.50.ht, 03.65.-w

## INTRODUCTION

Visualizations in multimedia resources and the use of analogical scaffolding can reduce cognitive load and help students learn abstract concepts [1]. The types of visual representations used can have a substantial impact on student thinking. They can facilitate the development of productive mental models, but may obstruct learning if the perceptual features are not aligned with the intended meaning [2, 3].

Single-photon experiments have played a vital role in understanding the significance of entanglement, and in driving the development of quantum information theory and technology [4]. Including them in an introductory course on quantum mechanics therefore presents an opportunity to simultaneously expose students to an exciting area of contemporary research, while developing quantum mechanical concepts via two-state systems, an approach that has been gaining favor in quantum mechanics instruction [5].

The use of research-based visualizations and simulations can help to improve students' understanding of the behavior of single photons in a Mach-Zehnder interferometer. Some physicists may object to any visual representation of a photon because it might reinforce classical ideas about quantum objects; however, prior education studies have shown that when instructors do not adequately attend to student models of quantum processes, these students will develop their own mental representations, which tend to be less expert-like [6]. Modeling is a key aspect of scientific reasoning, and can be used by students to develop a coherent picture of physical processes [7]. In the context of quantum phenomena, it is particularly important for students to recognize the bounds and limitations of a model, and that any visualizations of these processes should not lead to incorrect predictions for experimental outcomes.

We describe efforts to refine the visual representation of single-photon states in the QuVis interactive simulations [8], which aim to help introductory quantum physics students develop productive mental models of quantum processes. One of these simulations (*Interferometer Experiments*) compares and contrasts the behavior of classical particles, electromagnetic waves and single photons under the same experimental conditions. All of the single-photon simulations depict simplified, idealized situations (such as no background light and 100% efficient detectors), to reduce complexity and cognitive load. The simulations show each photon taking both paths of the interferometer after encountering a beam splitter, regardless of whether a second beam splitter is present in the apparatus. This is meant to be consistent with both the mathematical representation $1/\sqrt{2}$ ( |top path> + |bottom path> ) of the superposition state, and with delayed-choice experiments where the second beam splitter is inserted or removed while the photon is already in the apparatus [9].

## METHODOLOGY

An in-class trial of the *Interferometer Experiments* simulation in 2013 showed that the original visualiza-tion (see Fig. 1) led some students to develop incorrect ideas about quantum superposition; in particular, some students interpreted the effect of the beam splitter as being analogous to splitting a classical object into two pieces. In consultation with quantum optics researchers, through brainstorming and drawing on our own experience and the literature on student difficulties, we created four new animated sequences depicting a series of single photons incident on a beam splitter, the subsequent superposition state, and the detection of a photon in one of two detectors (shown as a flash in the detector). Each sequence was identical except for how the photon was represented.

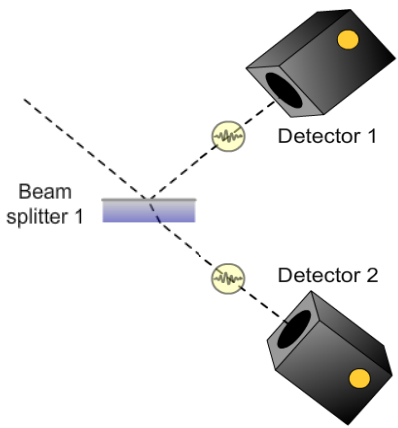

**FIGURE 1.** The original depiction of the single photon superposition state.

In the original depiction, the yellow coloring of the incident photon (chosen to suggest visible light) was opaque, then semi-transparent after encountering the beam splitter (Fig. 1), in order to provide a visual cue for the reduced probability amplitude associated with the individual state, and to reinforce that the state of the photon had changed. This general type of visualization was retained in the four revised depictions, but they also included more substantial cues about the differences between the incident and superposition states (see Fig. 2).

In each case, the amplitude of the incident wave is reduced by a factor of $1/\sqrt{2}$ for the superposition state, and the axis of the wave now rotates according to the direction of propagation. The wavelength remains constant throughout, to indicate that the energy of each component in the superposition is not halved (i.e., the wavelength is not doubled). We removed the outer black circle of the original depiction, which implied that each photon had a definite spatial extent; at the same time, the size of the yellow circle is the same for both incident and superposition states, to help students avoid the idea of the photon dividing like a classical object into two smaller components. The overall size of the photon has been increased in order to make all of these revised visual cues more perceptible.

As can be seen in Fig. 2, the yellow circle was removed entirely from one of the depictions (II), as a possible way to reduce associations with a classical particle. Visualizations III and IV use a dashed line connecting the two components of the superposition state in order to indicate that they remain linked while spatially separated, and are not independent of each other. Visualizations I and III are identical with the exception of the dashed line. The incident wave in Visualization IV is half red and half black, but then splits into two connected states, where one is entirely black and the other red.

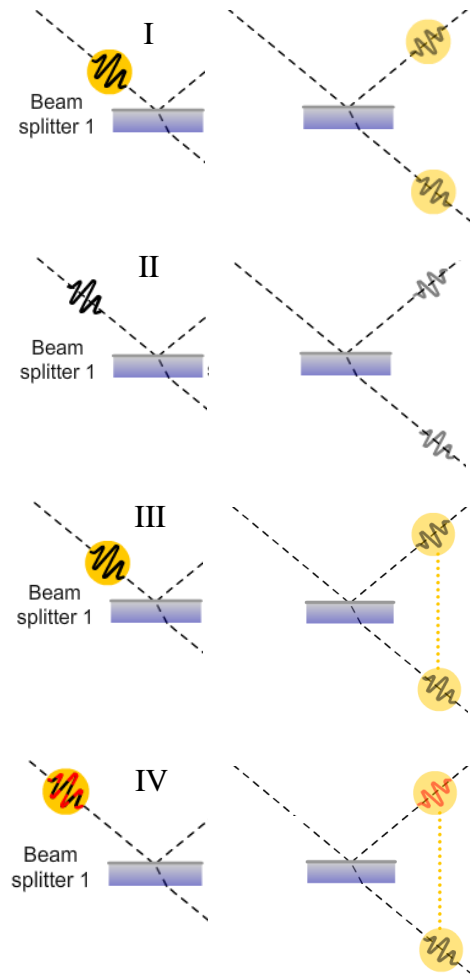

**FIGURE 2.** The four revised single photon depictions (numbered I to IV) used in the individual student interviews as described in the text. For each depiction, two screenshots are shown, one of the incident photon and one of the photon superposition state.

To assess the impact these revised visualizations might have on student thinking, we carried out individual interviews with nine volunteers who had just completed an introductory quantum physics course at the University of St Andrews. In these sessions, students were asked to view each of the revised visualizations in turn. They were first asked to freely describe what the visualization suggested to them, then given a list of statements (see Table 1) and asked to select out those that corresponded to what each visualization suggested to them. Students typically spent about 15 minutes in total exploring the different visualizations. The statements in Table 1 include incorrect student ideas about quantum superposition collected during the 2013 in-class trial of the *Interferometer Experiments* simulation (in particular, statements 2 & 4); incorrect ideas that we thought the new visualizations might suggest to students (statements 5 & 10); and ideas we considered productive for making sense of single-photon interference and delayed choice experiments (statements 1, 7, 8, 12 & 13).

**TABLE 1.** Statements used in the student interview sessions, in the same order as they were given.

| | |
|---|---|
| 1 | Photon takes both paths. |
| 2 | Photon has split into two photons. |
| 3 | Photon intensity is split in half. |
| 4 | Photon splits into two half-energy photons. |
| 5 | Two photons on top of one another have separated. |
| 6 | There are two photons that are entangled after the beam splitter. |
| 7 | Phase relationship between the two paths is maintained. |
| 8 | Photon is in a superposition state. |
| 9 | Photon is a wave packet. |
| 10 | Photon is a particle. |
| 11 | Photon is a wave packet and a particle at the same time. |
| 12 | Photon wave packet amplitude has decreased to conserve probability. |
| 13 | The two wave packets represent a single photon. |

## OUTCOMES

In these nine interviews, many students thought that Visualizations I and II were showing two distinct photons. Every student indicated that the dashed line joining the two components in Visualizations III and IV suggested some kind of relationship or connection was being maintained between them. Some students also agreed that the dashed line implied that the two photons remain in phase. Every student stated that Visualization II represented a wave packet, though several thought this made the photon seem too much like just a classical wave. Five of the nine students thought that Visualization IV showed two overlapping photons that then separate into two half-energy photons. The use of different colors in the two components seemed to suggest to them that they each have distinct properties (e.g., different energies or polarization states). Although the different colors were not meant to have any physical interpretation, other students were confused by what the two colors might actually represent. One commented that this particular depiction was overwhelming, because there was too much going on visually.

Figure 3 compiles results for two of the statements from Table 1. It was found that, in general, Visualization III most often suggested productive ideas about superposition states (with Visualization III chosen most often for statements 1, 7 and 13), while minimizing incorrect ideas (with Visualization III chosen least often for statements 2, 4 and 5). For the reasons outlined above, Visualizations I, II & IV were rejected, and Visualization III was adopted and incorporated into all of the QuVis single-photon simulations.

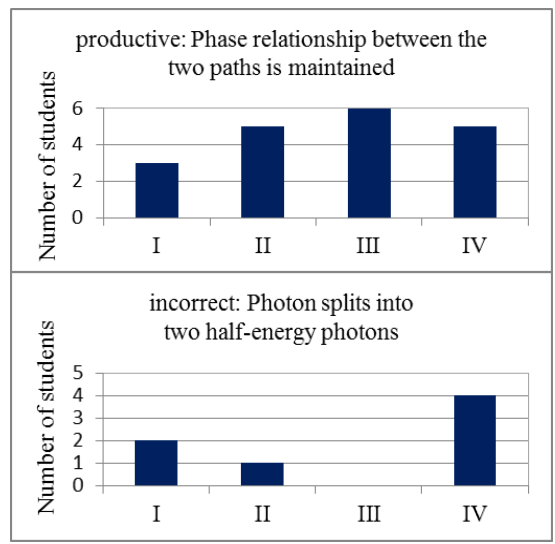

**FIGURE 3.** Outcomes for statements 7 (top) and 4 (bottom) of Table 1 for the different visualizations I to IV shown in Fig. 2.

In-class trials of the *Interferometer Experiments* simulation were carried out in two successive years during 50-minute computer workshop sessions; the original photon depiction was used in 2013, and Visualization III in 2014. Students in both years worked on similar activities during the session; for the 2014 session, the associated activity was shortened slightly, and some questions reworded. The implementations of the *Interferometer Experiments* simulation in 2013 and 2014 were somewhat different, with single-photon experiments in 2014 being discussed in a lecture prior to the simulation use. The 2013 course also discussed the single-photon Aspect experiments, but only after the workshop session.

At the end of the session students were asked to provide a written response (and to rate their confidence in their answer) to the question: "What happens when a single photon encounters a beam splitter?" [N=78 for 2014; N=28 for 2013, when only a a representative third of the class, namely those with a given lab day, used the simulation] We analyzed their responses with an emergent coding scheme, using the same codes for the 2013 and 2014 data. A subset of responses was coded by a second researcher and checked for inter-rater reliability. Categories with disagreement were discussed and revised until high inter-rater reliability was achieved (Cohen's Kappa 0.62 to 1).

Figure 4 shows results from our analysis of student responses collected during the in-class trials in 2013 and 2014; the various codes are given in the figure caption. One can see that the fraction of students describing a photon in a superposition state as being split into two separate entities (Category D in Fig. 4) decreased substantially from 2013 to 2014. An exact test for Pearson's chi-square shows a significant difference in the two distributions: $\chi^2(4, N=104) =15.9$; exact p=0.003. Some examples of typical student statements for the different codes are: [Category A] "The single photon will either go through or reflected, one way or the other.  It behaves like a particle."; [Category B] "It turns into a superposition going along both paths and only definitely goes down one of these paths when it is observed."; [Category C] "The single photon occupies two quantum states, of reflection and of transmission.  The photon does not 'split' but the probability of it being in either is equal."; [Category D] "A single photon is split into two after the beam splitter.  This is the photon exhibiting its wavelike properties."; [Category E] "It becomes superimposed and behaves much like a wave."

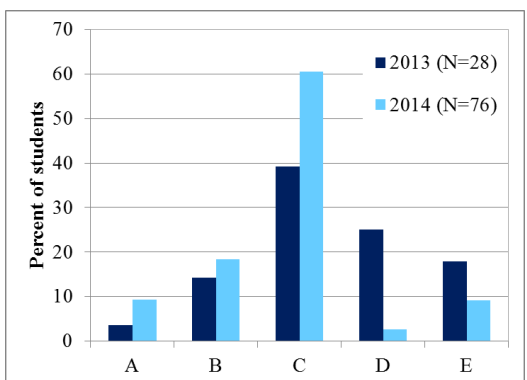

**FIGURE 4.** Results from the in-class trials as described in the text. The codes are: (A) "Photon takes one or the other path"; (B) "Measurement reveals the path the photon took"; (C) "Photon takes both paths"; (D) "Photon splits into two separate photons"; and (E) Other.

## CONCLUSIONS / FUTURE WORK

From the four visualizations we investigated in the interviews, Visualization III was optimal in terms of maximizing productive ideas about quantum superposition and minimizing incorrect ones. The class that used the revised visualization showed a lower frequency of incorrect ideas about quantum superposition. Limitations of this work are the small number of different visualizations assessed, and the somewhat different curricula in the 2013 and 2014 courses. Very few students noticed the change in amplitude of the wave packet in the transition from incident to superposition state, which may imply that this visual cue is too small to be easily seen. Other studies have found that students may incorrectly associate amplitude of a quantum wave with energy, so this visual cue may also not be productive [10].

We plan to continue these studies with further visualizations. We have also developed a number of visualizations of down-converted entangled photon pairs and are currently trialing these in student interviews.

## ACKNOWLEDGMENTS

We thank Frieder König at the University of St Andrews and Heather Lewandowski at JILA in Boulder, Colorado for useful discussions about single photons and their visualization. We gratefully acknowledge all of the students who participated in this study. We thank the UK Institute of Physics for funding this work.

## REFERENCES

1. N. Podolefsky and N. Finkelstein, *Phys Rev ST PER* **3**, 010109 (2007); Z. Chen, T. Stelzer and G. Gladding, *Phys Rev ST PER* **6**, 010108 (2010).
2. Z Chen and G Gladding, *Phys Rev ST PER* **10**, 010111 (2014).
3. A. Van Heuvelen and X. Zou, *Am. J. Phys.* **69**, 184 (2001).
4. A. Aspect, J. Bell and the second quantum revolution, in *Speakable and Unspeakable in Quantum Mechanics*, 2nd Ed. (Cambridge University Press, Cambridge, 2004).
5. Cf. M. Beck, *Quantum Mechanics – Theory and Experiment* (Oxford University Press, NY, 2012); D McIntyre, *Quantum Mechanics* (Pearson Addison-Wesley, San Francisco CA, 2012).
6. C. Baily and N. D. Finkelstein, *Phys Rev ST PER* **6**, 010101 (2010); *Phys Rev ST PER* **6**, 020113 (2010).
7. C. Schwarz et al., *J. Res. Sci. Teach.* **46**, 632 (2009).
8. Simulations 1-3 in the "New Quantum Curriculum sims" collection at www.st-andrews.ac.uk/physics/quvis
9. F. Kaiser, et al., *Science* **338**, 637 (2012).
10. S. B. McKagan, K. K. Perkins and C. E Wieman, *Phys Rev ST PER* **4**, 020103 (2008).